# „Schöne neue Lieferkettenwelt": Workers' Voice und Arbeitsstandards in Zeiten algorithmischer Vorhersage

Lukas Daniel Klausner, Maximilian Heimstädt, Leonhard Dobusch

## 1 Einleitung

Die Komplexität und zunehmend enge Kopplung vieler Lieferketten stellt eine große logistische Herausforderung für Leitunternehmen dar. Eine weitere Herausforderung besteht darin, dass Leitunternehmen – gedrängt durch Konsument:innen, eine kritische Öffentlichkeit und gesetzgeberische Maßnahmen wie die Lieferkettengesetze – stärker als bisher Verantwortung für Arbeitsstandards in ihren Zulieferbetrieben übernehmen müssen. In diesem Beitrag diskutieren wir einen neuen Ansatz, mit dem Leitunternehmen versuchen, diese Herausforderungen zu bearbeiten: die algorithmische Vorhersage von betriebswirtschaftlichen, aber auch ökologischen und sozialen Risiken.

Wir beschreiben die technischen und kulturellen Bedingungen für algorithmische Vorhersage und erklären, wie diese – aus Perspektive von Leitunternehmen – bei der Bearbeitung beider Herausforderungen hilft (Abschnitte 2–4). Anschließend entwickeln wir Szenarien, wie und mit welchen sozialen Konsequenzen algorithmische Vorhersage durch Leitunternehmen eingesetzt werden kann (Abschnitt 5). Aus den Szenarien leiten wir Handlungsoptionen für verschiedene Stakeholder-Gruppen ab, die dabei helfen sollen, algorithmische Vorhersage im Sinne einer Verbesserung von Arbeitsstandards und Workers' Voice weiterzuentwickeln (Abschnitt 6).

## 2 Management von Lieferketten – zwischen Komplexität und Verantwortung

Für Leitunternehmen in globalen Lieferketten stellen sich aktuell zwei unterschiedliche Herausforderungen. Die erste Herausforderung betrifft die zunehmende Komplexität vieler Lieferketten bei gleichzeitig enger Kopplung der einzelnen Glieder. Insbesondere von dieser Entwicklung betroffen sind Industrien und Branchen, in denen Leitunternehmen verschiedene Komponenten ihrer Endprodukte nicht von einigen wenigen Zulieferern, sondern von einer Vielzahl hochgradig spezialisierter Zulieferbetriebe beziehen (vgl. insbesondere die Typen der *relational* bzw. *captive value chains* nach Gereffi/Humphrey/Sturgeon 2005; siehe auch den Beitrag von Herr/Teipen/Gräf in diesem Band).

Zudem herrscht bei vielen Konsumgütern (beispielsweise Kleidung) ein erbitterter Preiskampf, sodass die Margen der Leitunternehmen vor allem mittels hoher Absatzzahlen und aufwendig organisierter Logistik erzielt werden („Fast Fashion"; Bhardwaj/Fairhurst 2010). Neben Versuchen, die Preise der einzelnen Komponenten niedrig zu halten, werden auch Transport und sonstige Logistik knapp kalkuliert und eng getaktet („just in time").

Aus dieser hohen Komplexität und engen Kopplung globaler Lieferketten ergibt sich für Leitunternehmen ein ausgeprägtes Bedürfnis nach möglichst genauen Informationen über Störungen im Betriebsablauf, insbesondere über Hindernisse im Fluss der Komponenten und Endprodukte aus Ländern des globalen Südens in Richtung der Heimatmärkte, die sich zumeist im globalen Norden befinden.

Aus der Frage, wie sich Lieferketten weiter rationalisieren lassen, ergibt sich für Leitunternehmen eine zweite Herausforderung: Verantwortung. Ambitionierte Renditeziele geben Leitunternehmen oftmals in Form von Preisdruck unmittelbar an ihre Zulieferbetriebe weiter. Beginnen globale Lieferketten in den Ländern des globalen Südens, stehen oft nur wenige institutionelle Barrieren zwischen diesem Preisdruck und den Sozialstandards für die Arbeiter:innen.

Der aus dem globalen Norden durchgereichte Preisdruck kann somit zu niedrigen Löhnen und mangelnden Sicherheitsmaßnahmen am Arbeitsplatz führen. Für längere Zeit schien es so, als würden Leitunternehmen diese Entwicklung nicht wahrnehmen oder sogar bewusst ignorieren oder leugnen. Die Motivation vieler Unternehmen zu stärkerer Verantwortungsübernahme für soziale sowie arbeits- und menschenrechtliche Standards in der Lieferkette schien ohne starken Druck von außen gering.

Dies hat sich in den letzten Jahren geändert, vor allem als Folge einer Reihe dramatischer Ereignisse – allen voran das Rana-Plaza-Desaster von 2013, bei dem beim Zusammensturz einer Kleidungsfabrik in Bangladesch mehr als 1000 Menschen ums Leben kamen (Schüßler/Frenkel/Wright 2019; Schüßler/Lohmeyer/Ashwin 2022). Durch die Berichterstattung rund um diesen und andere Unglücksfälle haben die Verbindungen zwischen Leitunternehmen im globalen Norden und ihren Zulieferbetrieben im globalen Süden mehr Aufmerksamkeit erfahren. Zunehmend müssen sich Leitunternehmen Fragen der Verantwortung für die sozialen Standards in ihren Zulieferbetrieben stellen.

Auch in diesem Zusammenhang steht die Hinwendung der Unternehmen zu Verfahren der algorithmischen Vorhersage von Unmut und sozialen Unruhen. Die Anbieter solcher informationstechnischer Verfahren versprechen den Unternehmen eine „schöne neue Lieferkettenwelt", in der sich die beiden Herausforderungen des Managements globaler Lieferketten – Verantwortung und Rationalisierung – gleichzeitig meistern lassen. Durch die Kombination aus öffentlich zugänglichen Daten (z. B. aus Social-Media-Plattformen) und lernfähigen Analyseverfahren soll es möglich werden, sowohl die Komplexität der globalen Wertschöpfung effizienter zu gestalten als auch unzureichende Arbeitsverhältnisse präziser und schneller sichtbar zu machen.

Es besteht jedoch auch die Gefahr, dass eine Beteiligung der Beschäftigten konterkariert wird, wenn die Verknüpfung personenbezogener digitaler Daten mit automatisierten Prognosen neue Formen der Überwachung ermöglicht; auch im Sinne einer antizipativen Repression. Das Ziel unseres Beitrages ist es daher, die Konsequenzen der algorithmischen Vorhersage einzuschätzen und daraus Empfehlungen für einen regulatorischen Rahmen abzuleiten.

Hierfür werfen wir in Abschnitt 3 zunächst einen Blick auf die Funktionsweise und Versprechen algorithmischer Vorhersageinstrumente. Anschließend beschreiben wir in Abschnitt 4, wie der Einsatz algorithmischer Vorhersage durch neue Lieferkettengesetze befördert wird. Anschließend entwickeln und analysieren wir in Abschnitt 5 drei Szenarien, wie algorithmische Vorhersage in Lieferketten mit Workers' Voice und Arbeitsstandards zusammenwirkt. Abschließend diskutieren wir in Abschnitt 6 regulatorische Konsequenzen und Wechselwirkungen mit weiteren Ansätzen zur Sicherstellung sozialer Standards in globalen Lieferketten.

## 3 Algorithmische Vorhersage in Lieferketten

Algorithmische Vorhersage bedient das Bedürfnis von Organisationen, Wissen über die Zukunft zu erlangen, um in der Gegenwart bessere Entscheidungen zu treffen. Algorithmische Vorhersage unterscheidet sich dabei von klassischen Formen des Zukunftswissens wie Szenarien oder Trends. Letztgenannte suggerieren Organisationen, dass sie ihre Handlungen an einem in der Zukunft liegenden Ereignis ausrichten können, ohne aber das Ereignis selbst beeinflussen zu können (Flyverbom/Garsten 2021).

Beispielsweise beschreibt ein Szenario, dass bei einem wahrscheinlichen Regierungswechsel in einem Zulieferland der Mindestlohn erhöht wird. Dieses Szenario erlaubt dem Leitunternehmen, nach neuen Lieferanten in anderen Ländern zu suchen. Das Unternehmen geht jedoch nicht davon aus, dass seine Suche den Ausgang der Wahl beeinflusst.

Dem Zukunftswissen der algorithmischen Vorhersage werden Fähigkeiten zugeschrieben, die weiter reichen. Bei algorithmischen Vorhersagen handelt es sich um sogenannte „operative Prognosen" (Singelnstein 2018): Vorhersagen, die sich auf Ereignisse in der nahen Zukunft beziehen und die den Ort der Entscheidung so schnell erreichen, dass das prognostizierte Ereignis durch Eingreifen in den Lauf der Dinge abgewendet oder zumindest abgeschwächt werden kann.

Die Figur der operativen Prognose lässt sich in verschiedenen gesellschaftlichen Bereichen antreffen. Im Polizeiwesen verspricht „Predictive Policing", dass sich mittels spezieller Algorithmen aus polizeiinternen und externen Daten ortsspezifische Einbruchswahrscheinlichkeiten vorhersagen lassen (Adensamer/Klausner 2021; Egbert/Heimstädt/Esposito 2021; zum verwandten Feld der „civil unrest prediction" vgl. Grill 2021). Im Versicherungswesen wird versprochen, dass Machine Learning mittels Daten über Kreditwürdigkeit und Fahrverhalten Vorhersagen über das individuelle Unfallrisiko treffen kann (Cevolini/Esposito 2020; Kiviat 2019).

Im Gesundheitswesen wiederum wird unter dem Begriff „Precision Medicine" versprochen, dass durch die algorithmische Auswertung biologischer, genetischer, historischer und sozioökonomischer Daten Krankheitsverläufe vorhergesagt und Präventions- und Behandlungsverfahren individualisiert werden können (Bíró et al. 2018). Folgt man den Angaben der Anbieter algorithmischer Vorhersage in Lieferketten, lässt sich die operative Prognose auf diesen Kontext übertragen – etwa durch Vorhersage eines nahenden Streiks, sodass ein Leitunternehmen Gespräche mit Arbeiter:innen vor Ort aufnehmen kann, um den Streik abzuwenden (Heimstädt/Dobusch 2021).

Wie aber wird dieses scheinbar so wirkmächtige Wissen über die Zukunft geschaffen? Wie auch andere Formen der operativen Prognose setzt sich algorithmische Vorhersage in Lieferketten aus zwei Komponenten zusammen: aus großen Datenmengen (Big Data) und maschinellem Lernen (Machine Learning).

„Big" werden die für eine Prognose genutzten Daten durch die Kombination verschiedener Quellen. Zum einen wird oft von eher „konventionellen" öffentlich zugänglichen Daten gesprochen, beispielsweise sozioökonomischen Daten zu Ländern und Märkten, aber auch Daten, die nur einen indirekten ökonomischen Bezug haben, wie etwa Wetterdaten. Zum anderen geben viele Anbieter algorithmischer Vorhersage an, auch Daten aus klassischen Nachrichtenquellen (etwa der Presse) und aus Social-Media-Plattformen zu kombinieren. Letzteres wird auch als „Open-Source Intelligence" (OSINT) bezeichnet.

Manche Social-Media-Unternehmen bieten über Schnittstellen direkten Zugriff auf ihre Daten, andernfalls lassen sich die Daten über sogenanntes Crawling oder Scraping verfügbar machen, d. h. automatisierte Verfahren, um Daten von Internetseiten abzurufen. Die Rolle von Machine Learning in der Auswertung dieser Daten liegt darin, die Informationen aus verschiedenen Sprachräumen zusammenzuführen und Muster über die unterschiedlichen Daten hinweg zu identifizieren. Hierzu werden sogenannte selbstlernende Algorithmen eingesetzt, die nach einer „Trainingsphase" (die oft durch Vorbereitung der Datensätze angeleitet, manchmal aber auch kaum bis gar nicht gesteuert ist) Muster in den Daten herausfiltern und darauf aufbauend Vorhersagen für die Zukunft treffen.

Für welche Ereignisarten genau solche algorithmischen Vorhersagen angestellt werden, unterscheidet sich je nach Anbieter; häufig werden jedoch politische Unruhen, Turbulenzen im Finanzsystem, Industrieunfälle oder ethisches Fehlverhalten im Management genannt. Mitunter schlüsseln Anbieter die Vorhersagen noch genauer auf. Beispielsweise beschreiben Heimstädt und Dobusch (2021, S. 202), dass ein Anbieter politische Unruhen weiter in „Proteste" und „Demonstrationen" unterteilt; Zwischenfälle im Bereich Corporate Social Responsibility (CSR) werden in „schlechte Arbeitsbedingungen" und „Menschenrechtsverletzungen" unterteilt.

Aus sozialwissenschaftlicher Sicht ist ein Blick in die Zukunft zwar nicht möglich (Luhmann 1976), aber die Art und Weise, wie sich Individuen und Organisationen auf die Zukunft beziehen, hat durchaus greifbare Konsequenzen (Beckert 2013; Mische 2009). Es lohnt sich daher zu fragen, wie und woher Verfahren wie die oben beschriebenen ihre kulturelle Plausibilität beziehen.

Warum glauben Entscheidungsträger:innen in Polizei, Versicherungswesen, Medizin oder Unternehmen daran, dass mit der Kombination von Big Data und Machine Learning der Blick in die Zukunft gelingen kann? Leonelli (2014) argumentiert, dass diese Kombination über die Zeit hinweg eine Aura der Allgemeingültigkeit bekommen habe, weil immer wieder sehr plausible, aber hochgradig kontextgebundene Beispiele verallgemeinert wurden.

Ein sehr bekanntes Beispiel hierfür ist die kund:innenspezifisch gezielte Produktwerbung der Supermarktkette *Target* in den USA: Das Kaufen gewisser Produkte korrelierte in den der Supermarktkette verfügbaren Daten damit, dass einige Monate später regelmäßig Wegwerfwindeln, Prä-Milch und andere Babyprodukte gekauft wurden, weshalb entsprechenden Kund:innen auf Basis ihrer vergangenen Einkäufe auf sie zugeschnittenes Werbematerial mit Schwangerschaftsprodukten zugesandt wurde. In einem Fall in Minneapolis bekam ein Haushalt auf diesem Wege Produktwerbung für die jugendliche Tochter zugeschickt, bevor die Eltern überhaupt von der Schwangerschaft wussten (Duhigg 2012).

Natürlich ist dieses Beispiel der Schwangerschaftsvorhersage durch Warenkorbdaten nicht mit der Vorhersage von Streiks in Lieferketten zu vergleichen. Bei genauerer Betrachtung scheint es sogar eher unwahrscheinlich, dass Softwareanbieter ohne jegliche Einbettung in eine lokale und situative

Arbeitsumgebung tatsächlich in der Lage sind, über unterschiedliche geografische Räume hinweg genaue Einschätzungen von Sozialstandards in Lieferketten zu leisten – und doch bietet eine wachsende Zahl von Unternehmen entsprechende Beratungsleistungen für Leitunternehmen an.

Ein Grund hierfür liegt unserer Ansicht nach in der *kulturellen* Legitimität algorithmischer Wirkmacht, die durch plakative Beispiele geschaffen und anschließend mit wenig „Reibungsverlusten" in „unwahrscheinlichere", d. h. weniger glaubhafte Anwendungsbereiche wie z. B. Lieferketten übertragen wurde. Eine weitere, nicht weniger wichtige Erklärung besteht in den aktuellen Gesetzgebungsvorhaben zu Lieferkettengesetzen, auf die Anbieter von algorithmischer Vorhersage sehr rasch reagiert haben.

## 4 Durch Algorithmen Lieferkettengesetze erfüllen?

Wie eingangs beschrieben scheint der Einsatz algorithmischer Vorhersage attraktiv für Leitunternehmen in Lieferketten, da das Verfahren verspricht, sowohl die Herausforderung der Komplexität als auch die der Verantwortungsübernahme bezüglich Arbeitsbedingungen zu bearbeiten.

In den letzten Jahren sind die Gesetzgeber in einigen Staaten des globalen Nordens zu dem Schluss gekommen, dass eine rein freiwillige Verantwortungsübernahme der Leitunternehmen nicht genügt, um das Problem schlechter Arbeitsbedingungen in Lieferketten zu beheben. Bemühungen, rechtliche Rahmenbedingungen für Leitunternehmen so zu verändern, dass sich die sozialen Standards in Zulieferbetrieben verbessern, wurden je nach Land von verschiedenen Akteur:innen vorangetrieben und haben bisher zu unterschiedlichen Ergebnissen geführt.

In einigen Ländern und Rechtsräumen wurden im Laufe der letzten zehn bis fünfzehn Jahre Sorgfaltspflichten entweder für spezifische Rechtsbereiche verabschiedet, z. B. 2019 in den Niederlanden (*Wet zorgplicht kinderarbeid* – Gesetz über eine Sorgfaltspflicht zur Vermeidung von Kinderarbeit) oder 2015 im Vereinigten Königreich (*Modern Slavery Act* – Gesetz gegen moderne Sklaverei), oder für bestimmte Branchen und Geschäftsfelder, z. B. 2021 in der Europäischen Union (Konfliktmineralien-Verordnung) oder 2010 in den

Vereinigten Staaten (*Conflict Minerals Provision*, formal Abschnitt 1502 des *Dodd Frank Act*).

Mit der deutschen Rechtslage sind hinsichtlich Umfang und Geltungsbereich der Sorgfaltspflichten am ehesten die Rechtslage in Frankreich (2017: *Loi relative au devoir de vigilance des sociétés mères et entreprises donneuses d'ordre* – Gesetz zu unternehmerischen Sorgfaltspflichten) und Norwegen vergleichbar (2022: *Åpenhetsloven* – Transparenzgesetz).

Über diese bereits verabschiedeten Gesetze hinaus gibt es noch zahlreiche weitere Gesetzesvorhaben und aus der Zivilgesellschaft kommende Entwürfe und Forderungen. Exemplarisch genannt seien hier nochmals die Europäische Union (Richtlinie zur Due-Diligence-Prüfung der Nachhaltigkeit von Unternehmen; erster Entwurf vorgelegt am 23. Februar 2022), Österreich („Initiative für ein Lieferkettengesetz") und die Schweiz (eidgenössische Volksinitiative „Für verantwortungsvolle Unternehmen – zum Schutz von Mensch und Umwelt", kurz Konzernverantwortungsinitiative; bei der Volksabstimmung am 29. November 2020 knapp gescheitert).

In Deutschland wurde das „Gesetz über die unternehmerischen Sorgfaltspflichten zur Vermeidung von Menschenrechtsverletzungen in Lieferketten" (kurz Lieferkettensorgfaltspflichtengesetz/LkSG) im Sommer 2021 beschlossen. Es trat zum 1. Januar 2023 zunächst nur für Unternehmen mit mindestens 3.000 Mitarbeiter:innen in Kraft; ab 1. Januar 2024 gilt es auch für alle Unternehmen mit mindestens 1.000 Mitarbeiter:innen.

Neben einigen allgemeineren Punkten (wie etwa Einrichtung eines Risikomanagements, Durchführung regelmäßiger Risikoanalysen, Einrichtung eines Beschwerdeverfahrens sowie Dokumentation und Berichterstattung) verpflichtet das Lieferkettengesetz die betroffenen Unternehmen insbesondere zur Umsetzung von Sorgfaltspflichten in Bezug auf Risiken bei Zulieferbetrieben. Bei den vom Lieferkettengesetz vorgesehenen Sorgfaltspflichten handelt es sich um sogenannte „Bemühenspflichten", d. h. die Unternehmen sind nicht zur Erbringung eines bestimmten Erfolgs verpflichtet, sondern müssen angemessene Vorkehrungen treffen, um eine Verletzung der Gesetzesbestimmungen zu verhindern.

Verantwortlich für die Prüfung der Unternehmen auf Befolgen und Einhalten des Gesetzes ist das Bundesamt für Wirtschaft und Ausfuhrkontrolle. Die Obergrenze des Strafausmaßes ist substanziell; verhängt werden können Strafen bis zu 800.000 Euro bzw. bei Unternehmen mit globalen Umsätzen von über 400 Millionen Euro auch bis zu 2 Prozent des globalen Umsatzes. Wird vom Bundesamt ein Bußgeld von mindestens 175.000 Euro ausgesprochen, können die so bestraften Unternehmen zusätzlich für bis zu drei Jahre für die Vergabe öffentlicher Aufträge gesperrt werden.

Für Leitunternehmen stellt sich somit die Frage, wie die aus den Lieferkettengesetzen neu erwachsenden Sorgfaltspflichten eingehalten werden können. Zu dieser Frage haben sich rasch eine ganze Anzahl freier Informationsangebote staatlicher, zivilgesellschaftlicher und privatwirtschaftlicher Stellen gebildet – staatlicherseits etwa vom Bundesministerium für wirtschaftliche Zusammenarbeit und Entwicklung (2022) und der Plattform „CSR in Deutschland" (2022), die vom Bundesministerium für Arbeit und Soziales betrieben wird. Von zivilgesellschaftlicher Seite aus ist z. B. die Initiative Lieferkettengesetz (2021) zu nennen, von privatwirtschaftlicher Seite der Bundesverband Materialwirtschaft, Einkauf und Logistik (2022) oder die Kanzlei CMS Deutschland (Bernhardt/Minderjahn/Wernecke 2021).

Darüber hinaus haben jedoch auch Anbieter algorithmischer Vorhersage die neuen Gesetze als Betätigungsfeld entdeckt. Es lässt sich beobachten, wie aktuell mehr und mehr Anbieter algorithmischer Vorhersage ihre Dienstleistungen zum „Dual-Use-Produkt" erklären, das sowohl der Rationalisierung und Effizienzsteigerung in den logistischen Abläufen dienen als auch bei der Erfüllung der gesetzlichen Sorgfaltspflichten und beim Bemühen um bessere Sozialstandards unterstützen soll.

Konkret im deutschen Markt ist die starke Vermarktung von algorithmischer Vorhersage als Werkzeug für das Lieferkettengesetz erkennbar; einzelne Anbieter offerieren spezielle Informationsveranstaltungen mit diesem Schwerpunkt, die auf ein gesteigertes Interesse seitens der Unternehmen stoßen. So erreichte ein von uns im Mai 2022 besuchter Online-Workshop Hunderte von Zuhörer:innen, die in der abschließenden Fragerunde vor allem Fragen zu rechtlichen Sachverhalten stellten, die teils sehr spezifisch waren.

# 5 Mögliche Folgen für Workers' Voice und Sozialstandards

Relevant sind nun weniger die technologischen Versprechungen algorithmischer Vorhersage, sondern vielmehr, wie und mit welchen Konsequenzen für Workers' Voice und Sozialstandards sie konkret in der Praxis von Lieferketten eingesetzt wird. Erweitern diese Vorhersagetechniken die Möglichkeiten von Arbeiter:innen, ihre Sichtweisen, Anliegen, Bedenken und Vorschläge in den betrieblichen Ablauf einzubringen? Oder erschweren sie es den Beschäftigten sogar, Gehör zu finden?

Die empirische Forschung zum Verhältnis von algorithmischer Vorhersage und Sozialstandards befindet sich aktuell noch am Anfang (vgl. die Forschungsagenda bei Heimstädt/Dobusch 2021). An dieser Stelle möchten wir daher ein Gedankenexperiment anstellen, um uns dieser Frage anhand von drei Szenarien zu nähern. Dabei sei vorausgesetzt, dass in der Zentrale eines Leitunternehmens im globalen Norden entscheidungsverantwortliche Führungskräfte durch algorithmische Vorhersage „vorgewarnt" werden, dass in einer Hafenanlage im globalen Süden ein Streik wegen Unzufriedenheit mit den Arbeitsbedingungen bevorstehen könnte.

Basis dieser Vorhersage ist, dass Hafenarbeiter:innen ihren Unmut über die Arbeitsbedingungen vor Ort vorab über Social-Media-Plattformen kundgetan haben. Diese Unmutsbekundungen wurden von Anbietern algorithmischer Vorhersage erfasst, ausgewertet und dem Leitunternehmen in Form einer Risikoprognose bereitgestellt. Grundlegend ergeben sich aus dieser Situation drei Handlungsoptionen für die Leitunternehmen:

- Im *1. Szenario* nutzt das Leitunternehmen die Vorhersage, um frühzeitig in den Dialog mit dem Zulieferunternehmen und/oder den Arbeitnehmer:innen vor Ort zu treten. So können Probleme oder Unstimmigkeiten thematisiert und ggf. ausgeräumt werden. Die Prognose ermöglicht es in diesem Szenario dem Leitunternehmen, die Arbeitsbedingungen vor Ort so schnell und genau zu verbessern, dass ein Streik abgewendet wird.
- Im *2. Szenario* nutzt das Leitunternehmen die Vorhersage, um frühzeitig ausweichende Maßnahmen im betroffenen Abschnitt der Lieferkette

einzuleiten (z. B. Nutzung alternativer Transportwege, anderer Zulieferunternehmen etc.). Das Leitunternehmen versucht in diesem Szenario nicht, auf die Ursachen des potenziellen Streiks einzuwirken, sondern nutzt die Vorhersage als Anlass, sich vorsorglich auf seine Folgen vorzubereiten.
- Im *3. Szenario* nutzen Leitunternehmen die Vorhersage, um Maßnahmen einzuleiten, die sich direkt gegen die Arbeitnehmer:innen oder ihre Vertretungen richten (z. B. Entlassung, Einschüchterung oder Bestechung). Wie im 1. Szenario haben auch die Maßnahmen in diesem Szenario das Ziel, den Ausbruch von Streiks und Protesten zu verhindern. Zum Einsatz kommen hierfür jedoch Mittel, die sich in eine lange Geschichte repressiver Maßnahmen in Arbeitskämpfen („Union Busting") einreihen.

Die arbeitssoziologische Forschung befasst sich schon seit einiger Zeit mit Fragen von Workers' Voice auf Social-Media-Plattformen (Heiland/Schaupp 2020; Hodder/Houghton 2015; Panagiotopoulos/Barnett 2015; Rosenblat 2018). Allerdings liegt der Fokus des Forschungsinteresses bislang vor allem auf den Strategien der Beschäftigten. Beispiele sind etwa gewerkschaftliche Kampagnenkommunikation (Bryson/Gomez/Willman 2010; Panagiotopoulos 2012), Mobilisierung trotz eines Mangels an institutionalisierter Interessenvertretung (Gerbaudo 2012; Tufekci 2017) oder die Selbstorganisation von Plattformarbeiter:innen (Wood/Lehdonvirta/Graham 2018).

Kaum bis gar nicht untersucht ist hingegen, wie Kommunikation und Inhalte auf diesen Social-Media-Plattformen von Dritten „beobachtet" und anderweitig verwendet bzw. verarbeitet werden. Zudem bleibt unklar, was eine solche Drittauswertung für die betroffenen Beschäftigten und ihre Kommunikation, Vernetzung und Selbstorganisation bedeutet.

Genau diese Drittbeobachtung ist aber das, was Anbieter algorithmischer Vorhersage versprechen: Als scheinbar unbeteiligte Dritte greifen sie wie oben beschrieben Daten zu arbeitsbezogenen Themen im großen Stil von Social-Media-Plattformen ab. Auf dieser Basis erstellen sie mithilfe ergänzender Daten Prognosen zu sozialen Prozessen und Ereignissen im Umfeld der Arbeitnehmer:innen, z. B. zu geplanten Streiks oder in Vorbereitung oder Durchführung befindlichen Gewerkschaftsgründungen.

Mit anderen Worten: Algorithmische Vorhersage greift auf Workers' Voice zurück, allerdings nicht als Mittel zur Sammlung, Weiterleitung und Verbreitung der Beschäftigtenbelange (siehe hierzu den Beitrag von Scheper/Vestena/Sorg/Zajak in diesem Band), sondern als äußere Beobachtung der auf Social-Media-Plattformen öffentlich einsehbaren Kommunikation über Arbeitsbedingungen, Vernetzung und Organisation der Beschäftigten.

Wie verhält sich nun algorithmische Vorhersage zu Workers' Voice? Die zuvor beschriebenen Szenarien liefern unterschiedliche Antworten auf diese Frage:

- Im *1. Szenario* leistet algorithmische Vorhersage einen Beitrag zu Workers' Voice. Wir können davon ausgehen, dass die auf Social-Media-Plattformen geteilten Unmutsbekundungen der Hafenarbeiter:innen sowohl konkrete Streikankündigungen als auch eher allgemeine Beschwerden über die Arbeitsbedingungen umfassen. Im Fall konkreter Ankündigungen werden diese über das System der algorithmischen Vorhersage schnell – vielleicht sogar schneller als über „klassische" Kanäle – an die Leitunternehmen und die Unternehmensleitung des Hafens übermittelt. Wenn diese rasch den Dialog mit den Arbeiter:innen aufnehmen, hat sich die algorithmische Vorhersage positiv auf Workers' Voice ausgewirkt.
- Im *2. Szenario* erreichen die Unmutsbekundungen zwar ebenfalls die Adressaten, jedoch ergeben sich hieraus keine positiven Konsequenzen für die Hafenarbeiter:innen; unter Umständen werden die Arbeitsbedingungen vor Ort sogar schlechter, wenn die Leitunternehmen als Reaktion auf die Streikvorhersage auf andere Logistikrouten umstellen. Am besagten Hafen würden somit als indirekte Folge Aufträge und schließlich Arbeitsplätze wegfallen. Die negativen Konsequenzen der Workers' Voice würden in diesem Szenario jedoch nicht von einzelnen Arbeiter:innen, sondern von der Gruppe getragen.
- Im *3. Szenario* hat algorithmische Vorhersage eindeutig negative Konsequenz auf Workers' Voice, denn in diesem Szenario werden die

algorithmischen Vorhersagen dazu genutzt, gegen einzelne Arbeiter:innen gezielt vorzugehen, mit der Rechtfertigung, dass „Gefahr im Verzug" bestehe.

Die Folge der algorithmischen Vorhersage im 3. Szenario ist als „prepression" zu verstehen. Dieser Portmanteau-Begriff, als Verschmelzung von „prevention" und „repression" von Schinkel (2011) geprägt, bezeichnet präventive Maßnahmen, die in ihrer konkreten Absicht und Ausgestaltung repressiven Charakter haben. Insbesondere im Zuge des Booms von „Predictive Policing" wurde dieser Analysezugang in den letzten Jahren immer relevanter, der an Science-Fiction-Erzählungen seit Mitte des 20. Jahrhunderts wie George Orwells „thoughtcrime" oder Philip K. Dicks „pre-crime" erinnert.

Wie dargelegt besteht auch im Kontext globaler Lieferketten die Gefahr „präpressiver" Maßnahmen als Antwort auf Organisierungs- und Arbeitskampfprozesse seitens der Arbeitnehmer:innen, was im Widerspruch zu den Zielen steht, die Worker-Voice-Ansätzen zugrunde liegen.

## 6 Diskussion und Ausblick

Algorithmische Vorhersage verspricht Leitunternehmen ein verbessertes Risikomanagement und die Möglichkeit, neue rechtliche Anforderungen und menschenrechtliche Sorgfaltspflichten zu erfüllen. Arbeitnehmervertretungen befürchten jedoch, dass die systematische Auswertung von Social-Media-Daten neue Möglichkeiten der Überwachung und Präpression von Arbeiter:innen bietet.

In diesem Beitrag haben wir erklärt, dass die Konsequenzen von algorithmischer Vorhersage für Arbeiter:innen nicht von der Technologie selbst abhängen – oder, mit Kranzberg (1986) gesprochen: „Technik ist weder gut noch böse; noch ist sie neutral." Vielmehr hängen die Konsequenzen von wechselseitig miteinander verschränkten Praktiken im Umgang mit algorithmischer Vorhersage in Lieferkettenkontexten ab, die sich grob entlang der Stakeholder-Kategorien Leitunternehmen, Softwareanbieter, Arbeitnehmervertretungen sowie Nichtregierungsorganisationen und kritische Öffentlichkeit systematisieren lassen.

**Leitunternehmen**

Anhand der drei Szenarien in Abschnitt 5 haben wir gezeigt, dass Leitunternehmen unterschiedliche Schlüsse aus algorithmischen Vorhersagen ziehen können. Manche dieser Schlüsse können Sozialstandards in Lieferketten zuträglich sein, andere lassen sich bereits in dieser frühen Phase der Technologieentwicklung kritisieren. Insbesondere die repressiven Negativszenarien sind aus menschenrechtlicher Sicht problematisch, aber auch die umgehenden Maßnahmen im „Graubereich-Szenario" werfen schon diverse Fragen von Verantwortlichkeit auf.

Sollte sich algorithmische Vorhersage dauerhaft in Leitunternehmen etablieren, ist damit zu rechnen, dass die Firmen symbolisch und strukturell auf Schwächen eingehen. Denkbar ist beispielsweise, dass Firmen einen menschen- und datenschutzrechtlich konformen Umgang mit Social-Media-Daten in bestehende Ethikkodizes aufnehmen. Denkbar scheint auch, dass Leitunternehmen die Frage des angemessenen Technologieeinsatzes an bestehende oder neue Berufsgruppen übertragen, beispielsweise „Corporate Digital Responsibility Officers" (Trittin-Ulbrich/Böckel 2022).

**Softwareanbieter**

Neue Lieferkettengesetze befördern die Nachfrage nach algorithmischer Vorhersage, wie in Abschnitt 4 beschrieben. Mit dem Wachstum der Branche werden jedoch auch viele Anbieter mit den beschriebenen Schwächen der Technologie konfrontiert. Sie reagieren darauf bereits mit der Veröffentlichung von Ethikkodizes, führen Begleitforschung zu sozialen Fragestellungen durch oder versprechen, ihre Dienste nicht ausschließlich Unternehmensführungen, sondern allen Stakeholder-Gruppen anzubieten, also auch der Arbeitnehmerseite.

Die meisten Anbieter algorithmischer Vorhersage schließen die in Abschnitt 5 beschriebene Präpression zwar offiziell aus, indem sie versichern, dass ihre Vorhersagen niemals personenbezogene Daten enthalten. Aus den Critical Data Studies wissen wir allerdings, dass ein solches Versprechen in dieser Absolutheit nicht zu halten ist – durch Zusatzwissen, Deanonymisierung und Ähnliches gibt es oftmals die Möglichkeit, aus allgemeinen Datenobjekten personenbezogene Rückschlüsse zu ziehen (vgl. z. B. Samarati 2001).

Denkbar scheint aktuell auch, dass Anbieter algorithmischer Vorhersage mittelfristig die Rahmung ihrer Dienstleistung verändern, um der Kritik auszuweichen. Beispielsweise könnte es im Kontext der Lieferkettengesetze für Softwareanbieter hilfreich sein, statt von „algorithmischer Vorhersage" von „algorithmischer Transparenz" zu sprechen und somit einen deutlich weniger kontroversen Begriff zu verwenden (Heimstädt/Dobusch 2020).

### Arbeitnehmervertretungen

Für organisationale, interorganisationale und transnationale Arbeitnehmervertretungen ergeben sich verschiedene, teilweise komplementäre strategische Optionen. Anschließend an in manchen Jurisdiktionen bestehende Mitwirkungsrechte von Arbeitnehmervertretungen beim Einsatz intraorganisationaler Datenverarbeitung (z. B. in Deutschland und Österreich) könnte die Etablierung vergleichbarer Informations- und Vetorechte auch für den Bereich interorganisationaler algorithmischer Vorhersagewerkzeuge in Lieferkettenkontexten ein Ansatz sein.

Eine Möglichkeit dafür wären globale Rahmenabkommen (siehe Kirsch/Puhl/Rosenbohm und Casey/Fiedler/Delaney in diesem Band). Unabhängig davon könnten Arbeitnehmervertretungen selbst algorithmische Vorhersage einsetzen, beispielsweise um Rekrutierungs- und Organisierungspotenziale zu identifizieren und damit Präpressionsgefahren zu begegnen.

Auf diese Weise könnten algorithmische Vorhersageinstrumente auch zur Verbesserung von Artikulationsbeziehungen im Lieferkettenmanagement beitragen. Eine Regulierung von algorithmischer Vorhersage, sei es in Form privatrechtlicher Vereinbarungen (etwa globaler Rahmenabkommen) oder nationaler bzw. internationaler Regulierungen, könnte durch die Ächtung bestimmter Nutzungsarten, die Einführung von Mindestanforderungen für Anbieter und den Einsatz von Vorhersagewerkzeugen den Boden für eine stärker kooperative Nutzung bereiten.

### Nichtregierungsorganisationen und kritische Öffentlichkeit

Die heutige Situation, in der Leitunternehmen zunehmend Verantwortung für Arbeitsstandards in ihrer Lieferkette übernehmen (müssen), ist zu einem beachtlichen Teil der Arbeit von Nichtregierungsorganisationen und einer kritischen Öffentlichkeit zu verdanken. Es scheint daher plausibel oder zumindest wünschenswert, dass diese auch die Auswirkungen algorithmischer Vorhersage kritisch begleiten.

Aktuell scheint noch unklar, ob und in welcher Weise dieses Thema von Nichtregierungsorganisationen oder Organisationen der digitalen Zivilgesellschaft (z. B. *Algorithm Watch*) aufgegriffen werden wird. Denkbare Positionierungen reichen von prinzipieller Opposition (z. B. Ächtung des Einsatzes derartiger Technologien) bis hin zur Mitwirkung bei der Ausgestaltung, z. B. durch Beratung von Gewerkschaften im Kontext globaler Rahmenabkommen.

Relevant für die zukünftige Entwicklung von algorithmischer Vorhersage ist auch die Wechselwirkung mit Worker Voice Tools (WVT). Diese neuen digitalen Werkzeuge sollen es Arbeiter:innen in globalen Lieferketten ermöglichen, Missstände am Arbeitsplatz kundzutun, auch wenn andere Kanäle nicht vorhanden sind (siehe Scheper/Vestena/Sorg/Zajak in diesem Band).

Bei der Konzeption und Weiterentwicklung solcher Werkzeuge scheint es sinnvoll, auch die zunehmenden Möglichkeiten algorithmischer Vorhersage zu berücksichtigen. Beispielsweise kann diese von Arbeitnehmervertretungen genutzt werden, um Kontexte zu ermitteln, in denen Organisierungsbedarf besteht, für den verstärkter Einsatz von Worker Voice Tools hilfreich sein könnte. Weiterhin könnte algorithmische Vorhersage auch in Worker Voice Tools integriert werden, beispielsweise um einzelne Unmutsbekundungen in diesen Tools mit allgemeinen Stimmungsbildern aus Social-Media-Daten zu kontextualisieren.

Abschließend lässt sich feststellen, dass die Auswirkungen von algorithmischer Vorhersage auf Arbeitsstandards und Workers' Voice nicht von der konkreten Technik bestimmt werden, sondern sich aus der Art der Nutzung durch Leitunternehmen und dem Einfluss von Stakeholdern in Lieferketten ergibt.

# Literaturverzeichnis

# Autoren

**Lukas Daniel Klausner** (https://l17r.eu) ist Mathematiker und Informatiker und forscht als Researcher an der Fachhochschule St. Pölten in den Bereichen Security, Privacy, Data Science und Science and Technology Studies. Sein Mathematikstudium an der TU Wien schloss er im Herbst 2018 mit einer Promotion *sub auspiciis* ab. Gemeinsam mit Paola Lopez hat er kürzlich den Arbeitskreis Mathematik trans- und interdisziplinär (AK MatriX, https://akmatrix.org) gegründet. Seine aktuellen Forschungsschwerpunkte sind Critical Algorithm and Data Studies, Ethik und Bias von Algorithmen, mathematische Grundlagen von Machine Learning/Artificial Intelligence und alle Fragestellungen, bei denen Technik und Gesellschaft aufeinandertreffen.

**Maximilian Heimstädt** ist Akademischer Oberrat an der Fakultät für Soziologie der Universität Bielefeld und Mitarbeiter im ERC-Projekt „The Future of Prediction". Zudem leitet er die Forschungsgruppe „Reorganisation von Wissenspraktiken" am Weizenbaum-Institut für die vernetzte Gesellschaft in Berlin. Der Schwerpunkt seiner Arbeit liegt auf Fragen der Organisations- und Managementforschung im Kontext der digitalen Transformation.

**Leonhard Dobusch**, Jurist und Betriebswirt, forscht und lehrt als Professor für Betriebswirtschaftslehre mit Schwerpunkt Organisation an der Universität Innsbruck sowie fungiert als wissenschaftlicher Leiter des Momentum Instituts in Wien. Seine Forschungsschwerpunkte liegen in den Bereichen organisationale Offenheit, Management digitaler Gemeinschaften sowie private Regulierung via Standards, vor allem im Bereich von Immaterialgüterrechten.